\pdfoutput=1
\documentclass[
  reprint,
  aps,
  pre,
  groupedaddress,
  amsmath,amssymb,
  floatfix
]{revtex4-2}

\usepackage{graphicx}
\usepackage{booktabs}
\usepackage{dcolumn}
\usepackage{bm}
\usepackage[dvipsnames]{xcolor}
\usepackage{balance}
\usepackage[T1]{fontenc}
\usepackage[utf8]{inputenc}
\usepackage{lmodern}
\usepackage{array}
\usepackage{float}

\usepackage[hidelinks]{hyperref}

\begin{document}

\title{Threshold-Controlled Geometric Reorganization in 2D Bootstrap Percolation}

\author{Fangfang Wang$^{1,5}$}

\author{Wei Liu$^{2,1}$}
\email{weiliu@xust.edu.cn}
\thanks{Corresponding author.}

\author{Kai Qi$^{4}$}

\author{Ying Tang$^{3}$}
\email{jamestang23@gmail.com}
\thanks{Corresponding author.}

\author{Zengru Di$^{1,5}$}

\affiliation{%
$^1$Department of Systems Science, Faculty of Arts and Sciences, Beijing Normal University, Zhuhai 519087, China \\
$^2$College of Science, Xi'an University of Science and Technology, Xi'an 710600, China \\
$^3$Institute of Fundamental and Frontier Sciences, University of Electronic Science and Technology of China, Chengdu 611731, China \\
$^4$2020 X-Lab, Shanghai Institute of Microsystems and Information Technology, Chinese Academy of Sciences, Shanghai 200050, China \\
$^{5}$International Academic Center of Complex Systems, Beijing Normal University, Zhuhai, China
}

\begin{abstract}
Two-dimensional bootstrap percolation is usually characterized by bulk observables, but whether increasing the activation threshold qualitatively reorganizes the geometry of the absorbing state has remained unclear. Here we show that the response undergoes a threshold-controlled geometric crossover. At low thresholds, the extrema of bulk and boundary-sensitive observables remain confined to a single collective low-$p$ window. At high thresholds, they split into distinct branches, revealing multiple geometric response scales. Over the accessible system sizes, the dominant finite-size signatures shift from fluctuations of the final active density to non-singleton boundary observables, while the fluctuation peak itself decreases. Time-resolved mechanism traces show that this crossover is accompanied by a progression from extended collective propagation to frontier exhaustion and, at the highest threshold, to quasi-one-step stabilization. Our results identify boundary organization as the dominant structural signature of high-threshold bootstrap percolation and show that conventional bulk observables alone do not capture the full reorganization of the absorbing state.
\end{abstract}  

\maketitle

\section{Introduction}

Bootstrap percolation is a prototypical irreversible activation process on lattices and networks, in which initially occupied sites evolve under deterministic threshold dynamics into an absorbing state \cite{Adler1991BootstrapPercolation,Hinrichsen2000Absorbing,Odor2004Universality}. Its local update rule and collective final outcome make it a standard model for spreading, jamming, and other constraint-driven cooperative phenomena in a wide range of complex systems \cite{Chalupa1979BootstrapPercolationBetheLattice,Aizenman1988MetastabilityBootstrapPercolation,Toninelli2006Jamming,Balogh2012SharpThresholdBootstrapPercolation,Balogh2005BootstrapPercolationRandomRegularGraph,Baxter2010BootstrapPercolationComplexNetworks,Gao2015BootstrapPercolationSpatialNetworks}. In two dimensions, the behavior depends strongly on the activation threshold: low thresholds promote extended propagation, whereas high thresholds sharply limit activatable sites and hence alter the path to the absorbing state \cite{Schonmann1992Bootstrap,Holroyd2003Sharp,Balogh2012SharpThresholdBootstrapPercolation,Ritort2003KCM,Berthier2011Glass}.

Most bootstrap-percolation studies have emphasized bulk observables, including the final active density, the probability of global activation, and the threshold dependence of the absorbing-state crossover \cite{Balogh2006Hypercube,Balogh2009BootstrapPercolationThreeDimensions}. While essential, these quantities do not fully capture the geometric organization of the absorbing state in a more detailed structural sense \cite{Saberi2015Recent,StaufferAharonyPercolation,Xue2024NucleationSpatialKCore}. States with similar active density can still differ substantially in interface content, cluster morphology, and the distribution of locally constrained motifs. This raises the structural question of how the absorbing-state geometry changes with increasing activation threshold beyond the bulk response.

This question is partly motivated by recent work on equilibrium lattice models, where geometry- and topology-sensitive observables have revealed signals beyond standard bulk thermodynamic measures \cite{Qi2018MIPAClassification,Wang2026Canonical,sitarachu2022evidence,DiCairano2024Topological,DiCairano2022Topological}. In the two-dimensional Ising model, isolated-spin and cluster observables capture higher-order signals that are less visible in density-like quantities \cite{Sitarachu2020ThirdOrderIsing,sitarachu2022evidence}. In the two-dimensional Potts model, boundary-sensitive observables were likewise found to be more informative than area-based measures for detecting structural changes in the disordered regime \cite{Liu2025PottsGeometry}. Bootstrap percolation is fundamentally different: its control parameter is the initial occupation probability rather than temperature, its dynamics are irreversible, and its evolution ends in absorbing states. The language of equilibrium transition orders therefore does not apply directly. Even so, these studies suggest a broader point relevant here: boundary and interface observables can encode structural information that bulk-density fluctuations alone may miss.

\begin{figure*}[t]
  \includegraphics[width=\textwidth]{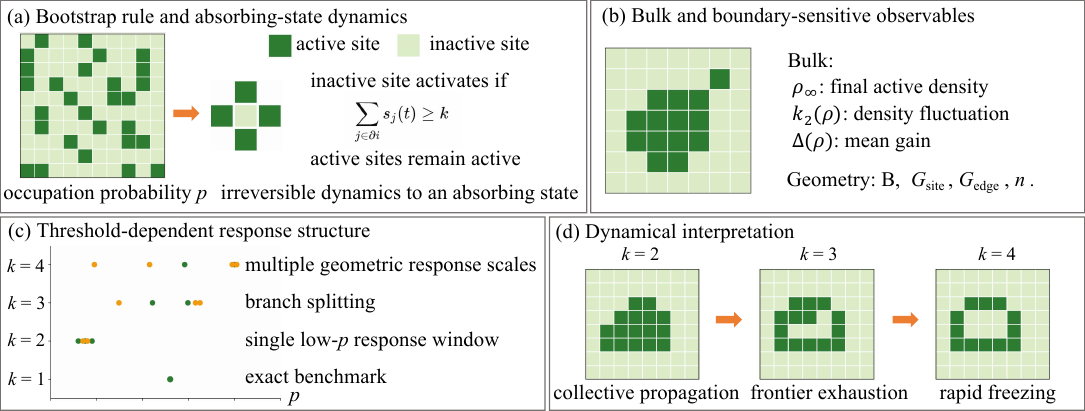}
  \caption{Overview of threshold-controlled geometric reorganization in two-dimensional bootstrap percolation. 
(\textbf{a}) Schematic of the bootstrap rule. Active sites remain active, while inactive sites become active if they have $\ge k$ active neighbors, leading to irreversible evolution toward absorbing states.  
(\textbf{b}) Bulk and boundary-sensitive observables. Bulk: final active density $\rho_\infty$, density fluctuation $k_2(\rho)$, and mean gain $\Delta(\rho)$. Boundary: edge density $B$, non-singleton boundary sites $G_{\mathrm{site}}^{\mathrm{ns}}$, non-singleton boundary edges $G_{\mathrm{edge}}^{\mathrm{ns}}$, and isolated-site signal $n$.
(\textbf{c}) Threshold-dependent response. Characteristic points show a single low-$p$ window for $k=2$, branch splitting for $k=3$, and multiple geometric scales for $k=4$, with $k_2(\rho)$ benchmark highlighted.  
(\textbf{d}) Dynamical interpretation. Snapshots for $k=2,3,4$ illustrate collective propagation, frontier exhaustion, and quasi-one-step freezing, highlighting boundary-dominated high-threshold absorbing states.
  }
  \label{fig:1}
\end{figure*}

We study how absorbing-state geometry reorganizes with increasing activation threshold in two-dimensional bootstrap percolation. Rather than focusing on the threshold itself or on bulk density alone, we examine how different observables encode the response hierarchy. Combining bulk and boundary-sensitive observables with time-resolved mechanism traces, we identify a threshold-controlled crossover: low thresholds exhibit a single collective low-$p$ response window, whereas high thresholds show a clear separation of characteristic response points. Over the accessible system sizes, the dominant finite-size signatures shift from bulk fluctuations to non-singleton boundary observables. The mechanism traces further reveal a progression from extended collective propagation to frontier exhaustion and finally to quasi-one-step geometric stabilization. These results identify boundary organization as the dominant structural signature of high-threshold bootstrap percolation; see Fig.~\ref{fig:1}.

\section{Model and Observables}

We consider two-dimensional bootstrap percolation on an $L \times L$ square lattice with periodic boundary conditions \cite{Adler1991BootstrapPercolation}. Each site is initially occupied independently with probability $p$, so that the initial state is specified by a binary variable $s_i(0) \in \{0,1\}$,
where $s_i(0)=1$ denotes an active site and $s_i(0)=0$ an inactive site. The total number of sites is $N=L \times L$.
Starting from the random initial condition, the system evolves according to the standard bootstrap rule with threshold $k$. At each synchronous update step, an inactive site becomes active if at least $k$ of its four nearest neighbors are active, while active sites remain active. Thus, if $s_i(t)$ denotes the state of site $i$ at update step $t$, the dynamics are
\begin{equation}
s_i(t+1)=
\begin{cases}
1, & s_i(t)=1,\\[4pt]
1, & s_i(t)=0 \ \text{and} \ \sum_{j \in \partial i} s_j(t) \ge k,\\[4pt]
0, & s_i(t)=0 \ \text{and} \ \sum_{j \in \partial i} s_j(t) < k,
\end{cases}
\label{eq:bp_rule}
\end{equation}
where $\partial i$ denotes the set of the four nearest neighbors of site $i$. The evolution continues until an absorbing state is reached $s_i(t+1)=s_i(t)$.

For each realization, we measure the final active density
\begin{equation}
\rho_{\infty}=\frac{1}{N}\sum_{i=1}^{N} s_i(\infty),
\end{equation}
where $s_i(\infty)$ denotes the absorbing-state configuration. To characterize the response of the system beyond the bulk density alone, we analyze several geometric observables extracted from the final configuration \cite{Katz1983Nonequilibrium}.

\subsection{Cluster observables}

Clusters are defined as connected components of active sites using nearest-neighbor connectivity on the periodic square lattice. For each active cluster, we measure its area and its boundary properties. In order to suppress the trivial contribution of isolated active sites to cluster geometry, we focus on non-singleton clusters when defining the perimeter-related observables.
We denote by $G_{\mathrm{site}}^{\mathrm{ns}}$ the average number of boundary sites in non-singleton clusters. A boundary site is an active site belonging to a non-singleton cluster that has at least one inactive nearest neighbor. Likewise, we denote by $G_{\mathrm{edge}}^{\mathrm{ns}}$ the average number of boundary edges in non-singleton clusters, where a boundary edge is an active--inactive nearest-neighbor edge along the cluster boundary.

Let the non-singleton clusters in a given absorbing-state configuration be labeled by $\ell=1,\dots,n_{\mathrm{ns}}$, and let $g_{\mathrm{site}}^{(\ell)}$ and $g_{\mathrm{edge}}^{(\ell)}$ denote the corresponding numbers of boundary sites and boundary edges of cluster $\ell$. Then
\begin{equation}
G_{\mathrm{site}}^{\mathrm{ns}}
=
\frac{1}{n_{\mathrm{ns}}}
\sum_{\ell=1}^{n_{\mathrm{ns}}}
g_{\mathrm{site}}^{(\ell)},
\end{equation}
and
\begin{equation}
G_{\mathrm{edge}}^{\mathrm{ns}}
=
\frac{1}{n_{\mathrm{ns}}}
\sum_{\ell=1}^{n_{\mathrm{ns}}}
g_{\mathrm{edge}}^{(\ell)}.
\end{equation}
In addition to cluster-wise averages, we also define the global boundary-edge density
\begin{equation}
B = \frac{N_{\mathrm{b}}}{N},
\end{equation}
where $N_{\mathrm{b}}$ is the total number of active--inactive nearest-neighbor edges in the absorbing state. The quantity $B$ measures the overall amount of interface in the final configuration and provides a global indicator of boundary complexity.

\subsection{Isolated-site signal and fluctuation observables}

To monitor the contribution of isolated active defects, we define the isolated-site signal $n$ as the fraction of active sites whose four nearest neighbors are all inactive. If $N_{\mathrm{iso}}$ denotes the number of such sites, then
\begin{equation}
n=\frac{N_{\mathrm{iso}}}{N}.
\end{equation}
Besides the geometric quantities, we also study the fluctuations of the final active density over independent realizations. Denoting by $\langle \cdots \rangle$ the average over realizations at fixed $p$, the second central moment is
\begin{equation}
k_2(\rho)
=
\left\langle
\left(\rho_{\infty}-\langle \rho_{\infty}\rangle\right)^2
\right\rangle .
\end{equation}
This quantity measures the sample-to-sample fluctuations of the absorbing-state density and serves as a bulk response indicator.
We further define the mean gain
\begin{equation}
\Delta \rho
=
\langle \rho_{\infty}-p \rangle ,
\end{equation}
which quantifies the net increase in the active density during the bootstrap process relative to the initially occupied fraction.

\subsection{Characteristic response points}

For each observable, we scanned its dependence on the initial occupation probability $p$ and defined a characteristic point $p^{*}$ as the location of a representative extremum. The resulting set of $p^{*}$ values provides a compact summary of how geometric and fluctuation observables depend on threshold and system size.
For each threshold $k$ and system size $L$, we generated $10^5$ independent random initial configurations at each sampled $p$ on an $L\times L$ square lattice with periodic boundary conditions, and evolved them by synchronous bootstrap updates to the absorbing state. We then computed $n$, $G_{\mathrm{site}}^{\mathrm{ns}}$, $G_{\mathrm{edge}}^{\mathrm{ns}}$, $B$, $k_2(\rho)$, and $\Delta \rho$, and averaged them over the realizations. The occupation probability was sampled on uniform grids with $N_T=121$ points: $L=45,60,75,90,120$ and $p\in[0,0.0008]$ for $k=1$; $L=30,60,90,120$ and $p\in[0.01,0.2]$ for $k=2$; and $L=60,90,120,180$ and $p\in[0.15,0.95]$ for $k=3$ and $k=4$.

To reduce small-scale numerical fluctuations, the response curves were smoothed by a moving average with an odd window size; the representative implementation used a window size of $9$. The characteristic points were then identified from the extrema of the smoothed curves and used as comparative markers of the response structure across observables, thresholds, and system sizes. In cases where no stable or physically meaningful extremum could be identified, the corresponding characteristic point was taken to be absent. Because these characteristic points are introduced as empirical descriptors of the response hierarchy, we do not interpret them as sharp critical points in the thermodynamic sense.

\begin{figure}[t]
  \includegraphics[width=\columnwidth]{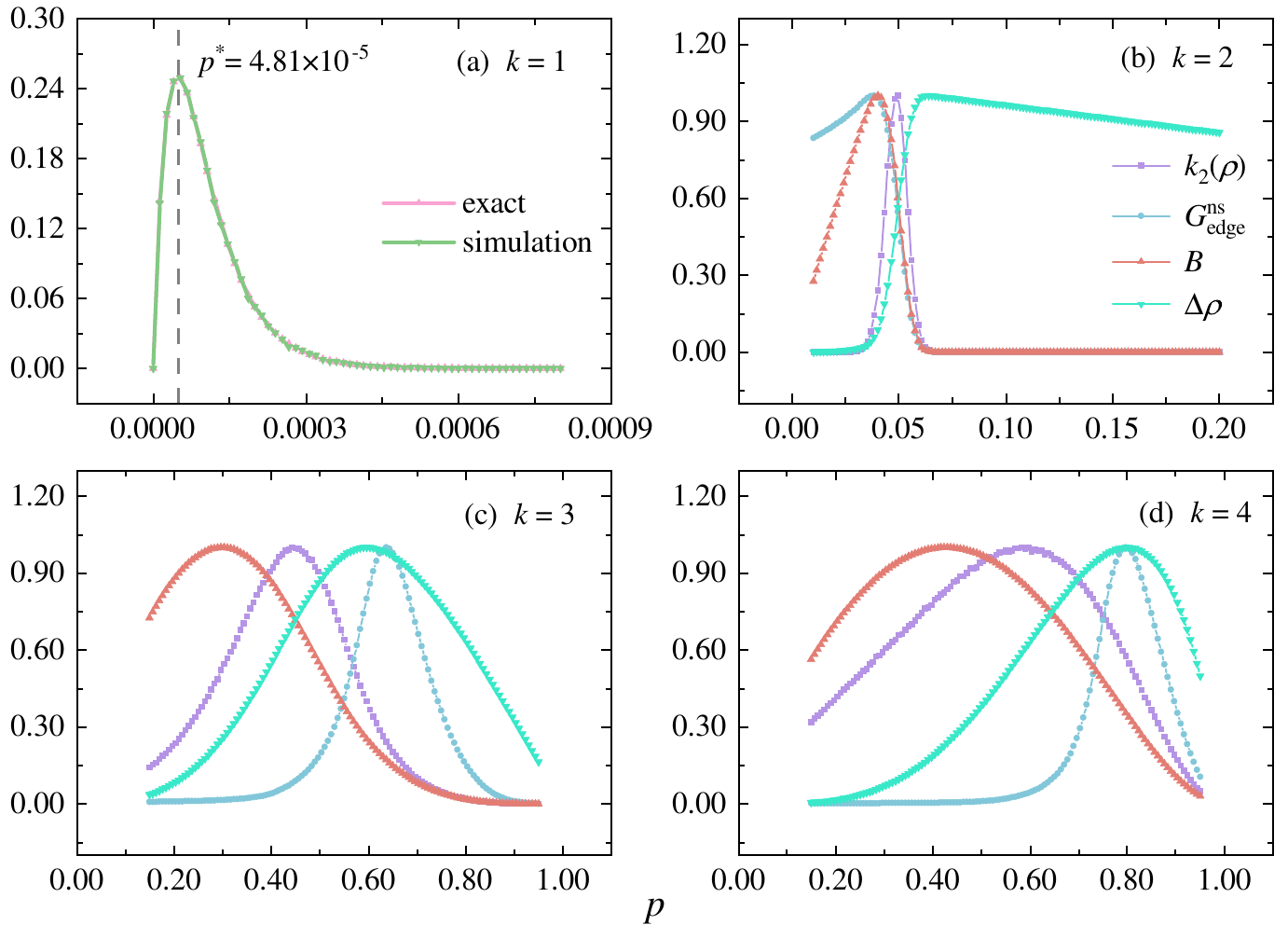}
  \caption{
  Normalized response curves of representative observables in two-dimensional bootstrap percolation for thresholds $k=1$--4. For $k=1$ [panel (a)], the peak of $k_2(\rho)$ matches the exact result, benchmarking the numerics. For $k=2$ [panel (b)], the responses of $k_2(\rho)$, $G_{\mathrm{edge}}^{\mathrm{ns}}$, $B$, and $\Delta \rho$ remain confined to a narrow low-$p$ interval, consistent with a single primary response window. For $k=3$ and $k=4$ [panels (c) and (d)], the peak positions separate clearly, indicating staged geometric reorganization of the absorbing state. Different observables reach their extrema at distinct $p$, showing that high-threshold dynamics are governed by multiple response scales rather than a single collective one.
  }
  \label{fig:response}
\end{figure}

\section{Results and Discussion}

We now examine how the absorbing-state response reorganizes with increasing threshold $k$. A clear crossover emerges: low thresholds exhibit a single collective low-$p$ response window, whereas high thresholds show separated geometric response scales and increasingly boundary-dominated finite-size signatures.

\subsection{The $k=1$ case: exact benchmark}

The case $k=1$ provides an exact benchmark for our numerical procedure. Here the dynamics is binary: any nonempty initial configuration activates the entire lattice, whereas the empty configuration remains absorbing. Accordingly, only the fluctuation measure $k_2(\rho)$ exhibits a robust extremum, while the geometric observables do not.
As shown in Fig.~\ref{fig:response}(a), the numerical peak position agrees closely with the exact result, and the peak height approaches the expected value $k_2^{*}=1/4$. This agreement validates the simulation protocol and the smoothing-based extremum identification used throughout the paper, and provides a clean reference for the more structured behavior at higher thresholds.

\subsection{The $k=2$ case: a narrow primary response window}

For $k=2$, the characteristic extrema of the absorbing-state observables remain confined to a narrow low-$p$ interval. The peaks of $n$, $B$, $G_{\mathrm{site}}^{\mathrm{ns}}$, $G_{\mathrm{edge}}^{\mathrm{ns}}$, $k_2(\rho)$, and $\Delta \rho$ all lie in the same general region and shift together toward smaller $p$ with increasing system size; see Table~\ref{tab:peak_positions} and Fig.~\ref{fig:pc}(b). This coordinated drift indicates that the response is still controlled by a single low-$p$ activation regime rather than by several separated geometric scales.

\begin{figure*}[t]
  \includegraphics[width=0.8\textwidth]{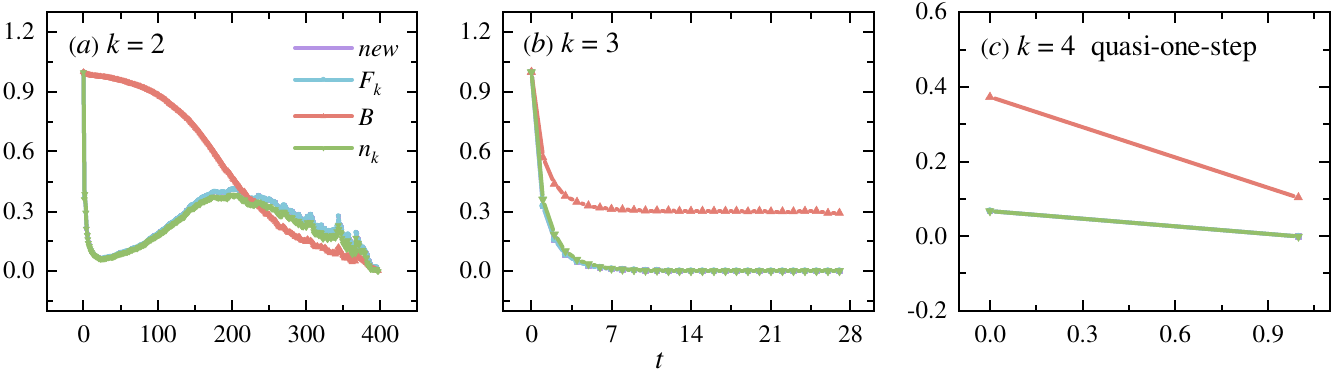}
  \vspace{-1em}
  \caption{
Normalized mechanism traces for representative thresholds in two-dimensional bootstrap percolation. Shown are the normalized time traces of the newly activated fraction ($\mathrm{new}$), the activatable-front density ($F_k$), the boundary-edge density ($B$), and the fraction of inactive sites with exactly $k$ active neighbors ($n_k$). Panel (a) [$k=2$] shows extended collective propagation. Panel (b) [$k=3$] shows rapid depletion of the activatable front and newly activated fraction, while $B$ approaches a finite plateau, indicating frontier exhaustion. Panel (c) [$k=4$] exhibits quasi-one-step dynamics, followed by rapid freezing into an absorbing state set by stable vacancy clusters. Together, the traces reveal a threshold-dependent progression from collective propagation to frontier exhaustion and finally to quasi-one-step geometric stabilization.
}
  \label{fig:mechanism_trace}
\end{figure*}

Although these observables probe different aspects of the absorbing-state structure, their characteristic points remain strongly co-localized at $k=2$. Combined with the extended collective propagation seen in Fig.~\ref{fig:mechanism_trace}(a), this behavior supports the interpretation that the absorbing-state response is still organized by a single underlying collective propagation process. The $k=2$ case is therefore best viewed as a collective low-$p$ regime that precedes the clearer branch separation and boundary-dominated behavior found at higher thresholds.

\begin{figure}[t]
  \includegraphics[width=\columnwidth]{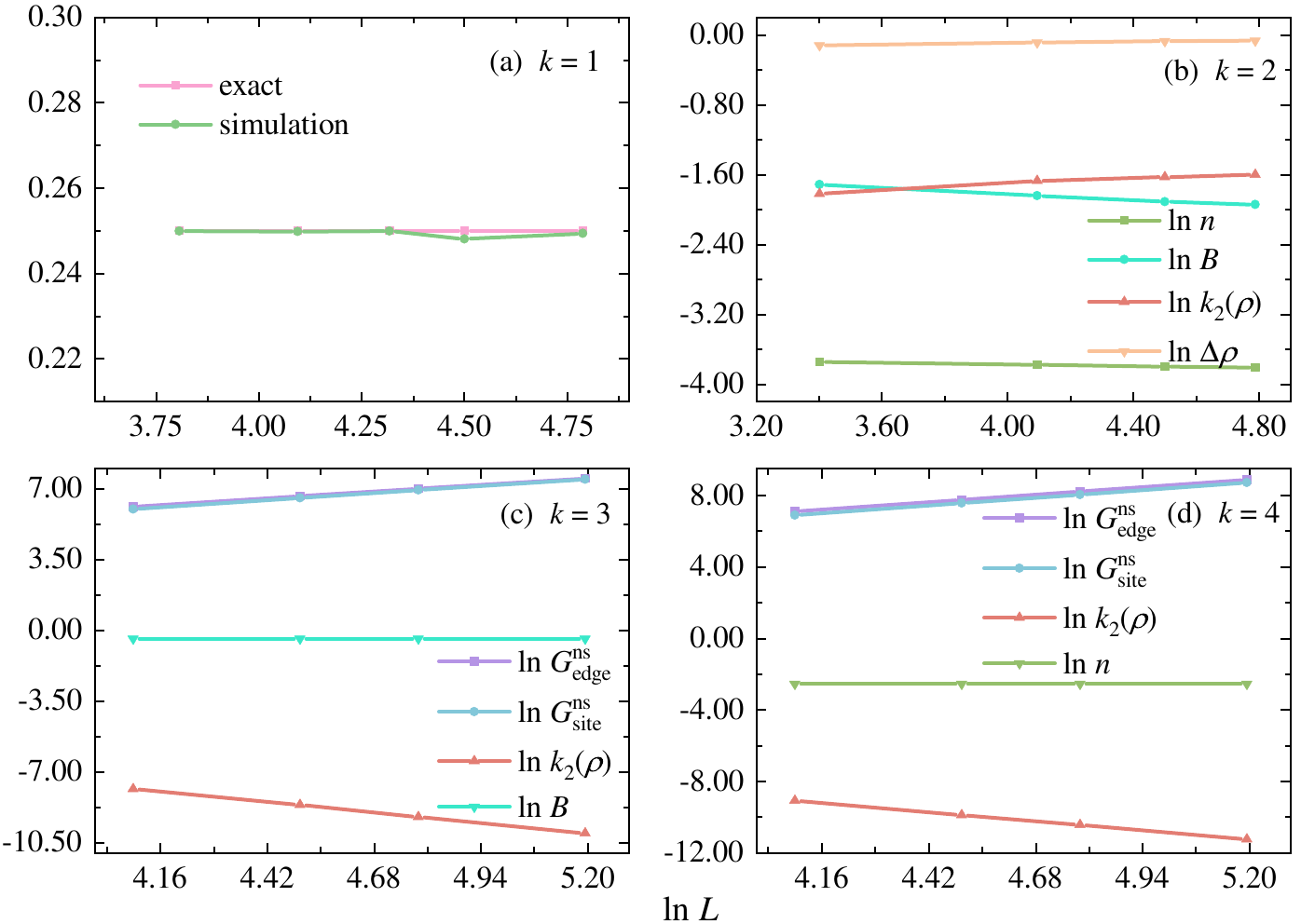}
 \caption{
Logarithmic plots of the peak amplitudes of observables in two-dimensional bootstrap percolation for thresholds $k=1$--4. For $k=1$ [panel (a)], the peak height of $k_2(\rho)$ remains close to the exact binary-response value. For $k=2$ [panel (b)], the peak amplitudes show only weak size dependence. For $k=3$ and $k=4$ [panels (c) and (d)], the peak amplitudes of $G_{\mathrm{site}}^{\mathrm{ns}}$ and $G_{\mathrm{edge}}^{\mathrm{ns}}$ increase strongly over the available size range, whereas that of $k_2(\rho)$ decreases. This contrast indicates that the dominant finite-size signatures shift toward boundary-sensitive observables at high thresholds.
}
  \label{fig:ln}
\end{figure}

\subsection{High-threshold response: $k=3$ and $k=4$}

\begin{figure}[t]
  \includegraphics[width=\columnwidth]{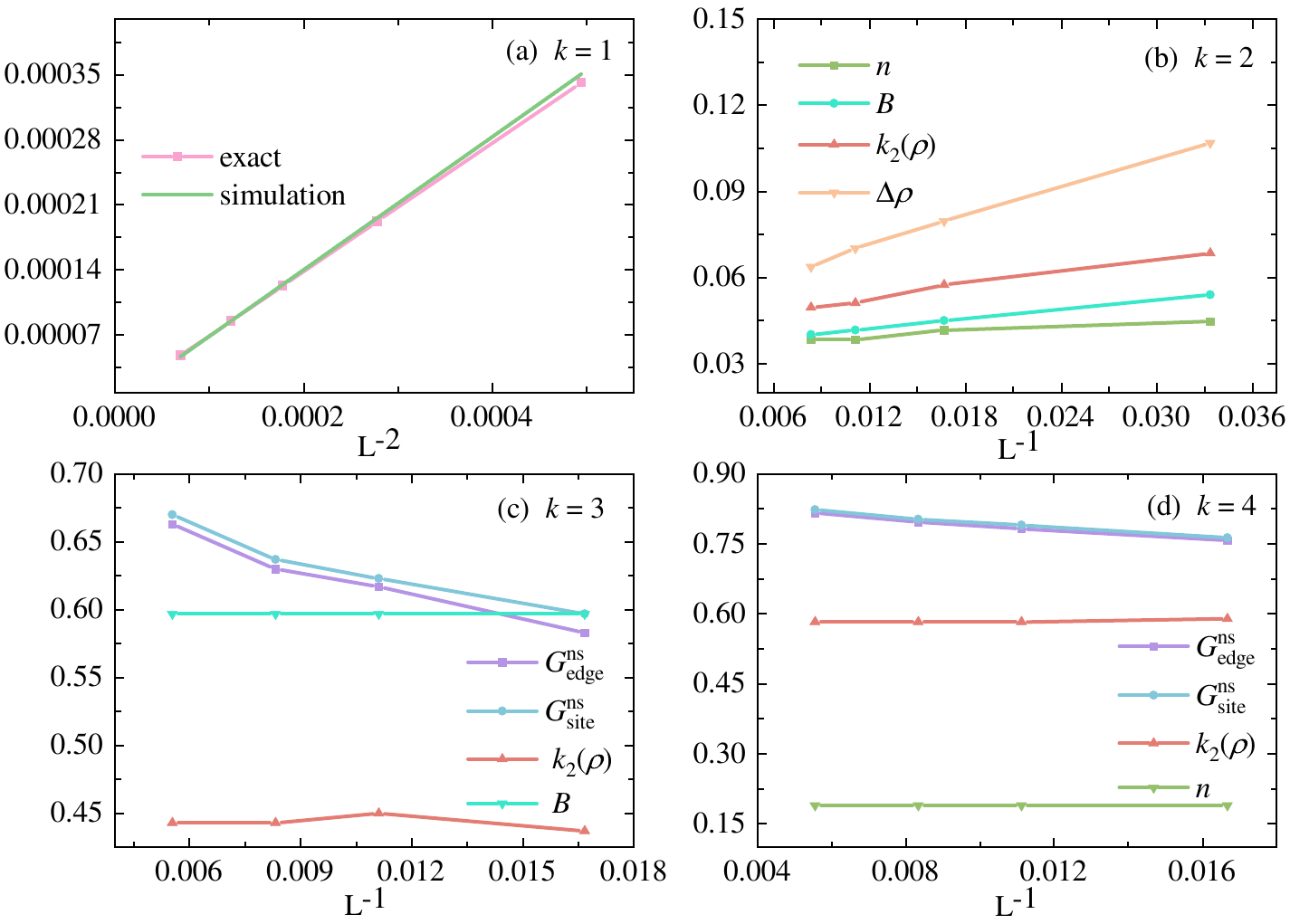}
  \vspace{-0.5em}
  \caption{
  System-size dependence of the characteristic positions $p^{*}$ for different observables in two-dimensional bootstrap percolation at thresholds $k=1$--4. For $k=1$ [panel (a)], the numerical peak position of $k_2(\rho)$ matches the exact result. For $k=2$ [panel (b)], the characteristic positions of $n$, $B$, $k_2(\rho)$, and $\Delta \rho$ remain close and shift together toward smaller $p$ with increasing $L$, consistent with a single low-$p$ response window. For $k=3$ and $k=4$ [panels (c) and (d)], the characteristic positions split into distinct branches, indicating multiple geometric response scales at high thresholds.
  }
  \label{fig:pc}
\end{figure}

A qualitative reorganization of the absorbing-state response emerges for $k=3$ and $k=4$. In both cases, the characteristic extrema no longer cluster within a single low-$p$ window but separate systematically along the $p$ axis. The boundary-edge density $B$ reaches its extremum at relatively small $p$, the fluctuation peak of $k_2(\rho)$ appears at larger $p$, and the non-singleton boundary observables $G_{\mathrm{edge}}^{\mathrm{ns}}$ and $G_{\mathrm{site}}^{\mathrm{ns}}$, together with the mean gain $\Delta \rho$, peak further to the right. This branch separation is evident in Fig.~\ref{fig:response}(c,d) and Fig.~\ref{fig:pc}(c,d). The isolated-site signal $n$ is not universal in this regime: it is absent for $k=3$ but shows a clear extremum for $k=4$.

This ordered separation shows that high-threshold bootstrap percolation is no longer governed by a single dominant response scale. Instead, as $p$ increases, the system passes through distinct geometric response stages: maximal boundary complexity, then maximal sample-to-sample fluctuation in the final active density, and finally the strongest boundary signatures of extended non-singleton structures. The finite-size dependence supports this picture. As shown in Fig.~\ref{fig:ln}(c,d), the peaks of $G_{\mathrm{site}}^{\mathrm{ns}}$ and $G_{\mathrm{edge}}^{\mathrm{ns}}$ grow strongly with $L$, whereas the peak of $k_2(\rho)$ decreases; by contrast, the peak of $B$ varies only weakly, and that of $n$ is nearly size independent for $k=4$. Thus, in the high-threshold regime, the finite-size signatures are carried increasingly by boundary organization rather than by fluctuations of the active density alone.

\vspace{-0.5em}
\begin{table*}[htbp]
\centering
\caption{Finite-size summary of characteristic response observables in two-dimensional bootstrap percolation.}
\label{tab:summary_scaling}
\begin{tabular}{lll}
\toprule
Regime & Peak-position structure & Finite-size tendency \\
\midrule
$k=1$
& $p_{k_2}^{*}\sim L^{-2}$
& $k_2^{*}\rightarrow 1/4$ \\
\addlinespace[3pt]
$k=2$
& different $p_a^{*}$ remain in one narrow low-$p$ window
& no strong growth in boundary observables \\

$k=3$
& characteristic positions split into separated branches
& \shortstack[l]{over the available sizes, $A_{G_{\mathrm{site}}^{\mathrm{ns}}}^{*}$ and $A_{G_{\mathrm{edge}}^{\mathrm{ns}}}^{*}$ grow strongly, \\
while $A_{k_2}^{*}$ decreases} \\

$k=4$
& characteristic positions split into separated branches
& \shortstack[l]{over the available sizes, $A_{G_{\mathrm{site}}^{\mathrm{ns}}}^{*}$ and $A_{G_{\mathrm{edge}}^{\mathrm{ns}}}^{*}$ grow strongly, \\
while $A_{k_2}^{*}$ decreases} \\
\bottomrule
\end{tabular}
\end{table*}

\subsection{Finite-size signatures of the response hierarchy}

The finite-size data further clarify that the response structure of two-dimensional bootstrap percolation depends qualitatively on the threshold $k$. Here we focus on the size dependence of the characteristic extrema extracted from the absorbing-state observables.

For a generic observable $O_\alpha(L,p)$, we denote by $p_\alpha^*(L)$ the characteristic position of its response curve and by
\begin{equation}
A_\alpha^*(L)=O_\alpha\!\left(L,p_\alpha^*(L)\right)
\end{equation}
the corresponding peak amplitude. 
Over the available system sizes, the peak amplitudes can be summarized by effective size-dependent trends. For compactness, we parameterize these trends by
\begin{equation}
A_\alpha^*(L)\propto L^{y_\alpha^{\mathrm{eff}}},
\label{eq:peak_scaling_short}
\end{equation}
where $y_\alpha^{\mathrm{eff}}$ provides only a local summary of the observed size dependence over the simulated range. Because the present analysis is based on a limited set of system sizes and on characteristic points extracted from smoothed response curves, we do not interpret these exponents as asymptotic critical exponents.

For $k=1$, the system provides an exact benchmark. In this case, the final density is binary, $\rho_\infty\in\{0,1\}$, and
\begin{equation}
k_2(\rho)=Q_L(p)\bigl[1-Q_L(p)\bigr],
\end{equation}
\begin{equation}
Q_L(p)=1-(1-p)^{L^2}.
\end{equation}
The peak condition $Q_L=1/2$ gives
\begin{equation}
p_{k_2}^*
=
1-2^{-1/L^2}
=
\frac{\ln 2}{L^2}+O(L^{-4}),
\end{equation}
with the exact peak height $k_2^*=1/4$. The numerical data are consistent with this result.

For $k=2$, the characteristic extrema of different observables remain confined to a narrow low-$p$ interval and drift together as $L$ increases. This behavior supports the interpretation of a single primary response window rather than several separated geometric response scales. Correspondingly, the peak amplitudes show no strong growth of the boundary observables.
A qualitatively different finite-size structure emerges for $k=3$ and $k=4$. In these cases, the characteristic positions no longer collapse into one narrow window, but instead remain separated over the accessible system sizes. At the same time, different observables show clearly different size dependences.

To summarize these trends more clearly, we collect the finite-size results in Table~\ref{tab:summary_scaling}. The four thresholds fall into three regimes. The case $k=1$ provides an exact benchmark with $p_{k_2}^{*}\sim L^{-2}$ and $k_2^{*}=1/4$. For $k=2$, the characteristic extrema remain confined to a single low-$p$ response window. By contrast, for $k=3$ and $k=4$, the characteristic positions split into separated branches, while the dominant finite-size signatures are carried by $G_{\mathrm{site}}^{\mathrm{ns}}$ and $G_{\mathrm{edge}}^{\mathrm{ns}}$ rather than by $k_2(\rho)$. Over the available system sizes, these trends provide direct support for the crossover from collective low-$p$ response to boundary-controlled geometric stabilization. Detailed numerical values of the characteristic peak positions and peak amplitudes are provided in Appendix A (Tables~\ref{tab:peak_positions} and \ref{tab:peak_amplitudes}).

To clarify the dynamical origin of the threshold-dependent response hierarchy, we tracked the time evolution of the newly activated fraction, the activatable-front density $F_k$, the boundary-edge density $B$, and the fraction of inactive sites with exactly $k$ active neighbors. In Fig.~\ref{fig:mechanism_trace}(a), the $k=2$ case shows extended collective propagation, consistent with a single low-$p$ response window. For $k=3$ [Fig.~\ref{fig:mechanism_trace}(b)], the activatable front is rapidly depleted and the newly activated fraction falls to zero, while the boundary signal remains finite. The arrest therefore results from the disappearance of frontier sites rather than the loss of boundary structure. For $k=4$ [Fig.~\ref{fig:mechanism_trace}(c)], the dynamics is essentially quasi-one-step, reflecting the highly restrictive requirement that all four nearest neighbors be active. This directly supports the crossover from low-threshold collective response to high-threshold boundary-controlled geometric stabilization.

Taken together with the response-curve and finite-size analyses, the mechanism traces show that the high-threshold regime is increasingly controlled by activatable boundary geometry rather than the bulk active fraction alone. For $k=3$, propagation stops because the activatable front is exhausted, although substantial interface structure remains in the absorbing state. For $k=4$, the dynamics becomes essentially quasi-one-step, indicating that extended propagation is no longer sustained and the system rapidly freezes into a geometry-selected absorbing configuration. These dynamical signatures provide a natural interpretation of why boundary-sensitive observables are especially informative relative to bulk-density fluctuations in the high-threshold regime.

\subsection{Relation to equilibrium geometric observables}

The present results are comparable to earlier geometric studies of equilibrium lattice systems only at the level of structural observables, not transition order. In the two-dimensional Ising and Potts models, boundary-sensitive observables were found to resolve structural changes more clearly than bulk- or area-like quantities. The common lesson relevant here is therefore methodological rather than thermodynamic: different geometric observables are not equally informative, and boundary-related quantities can reveal structural reorganization that bulk measures partly obscure.

Bootstrap percolation differs fundamentally from these equilibrium systems. Its control parameter is the initial occupation probability $p$, the dynamics are irreversible, and the evolution terminates in absorbing states. Moreover, there is no underlying microcanonical entropy or density of states from which an inflection-point analysis could be constructed. The present results should therefore not be interpreted as evidence for an Ising- or Potts-type higher-order transition in the strict thermodynamic sense. What our data do show is a threshold-dependent reorganization of absorbing-state geometry, together with a shift from bulk-density fluctuations to boundary-sensitive observables as the clearest indicators of that reorganization. In that sense, the connection to equilibrium geometric studies is heuristic but useful.

\section{Conclusion}

We have shown that two-dimensional bootstrap percolation exhibits a threshold-controlled crossover in the geometry of its absorbing states. For low thresholds, the characteristic extrema of different observables remain confined to a single low-$p$ response window, consistent with a collective primary response. For high thresholds, these characteristic points separate into distinct branches, revealing a hierarchy of geometric response scales. Over the accessible system sizes, the strongest finite-size trends are carried by the non-singleton boundary observables rather than by the fluctuation measure $k_2(\rho)$. Time-resolved mechanism traces further show that this crossover is accompanied by a progression from extended collective propagation to frontier exhaustion and finally to quasi-one-step geometric stabilization.

These results do not imply an equilibrium-style higher-order transition scenario. They do show, however, that high-threshold bootstrap percolation is increasingly governed by boundary organization rather than by bulk-density fluctuations alone. Boundary-sensitive observables therefore provide the clearest probes of absorbing-state reorganization in this irreversible threshold system. More broadly, the present results suggest that, in cooperative activation processes with sufficiently restrictive local support, the macroscopic response may become increasingly shaped by activatable boundary motifs rather than by bulk-density fluctuations alone.

\section*{ACKNOWLEDGMENTS}
\vspace{-8pt}
This work was supported by China’s National Natural Science Foundation Grant Nos. 12575033, 12304257, 12322501 and 12575035. Y.T. acknowledges support from the Natural Science Foundation of Sichuan Province
(Grant No. 2026NSFSCZY0124). Computing resources were provided by the Interdisciplinary Intelligence
Supercomputing Center of Beijing Normal University, Zhuhai.

\balance
\bibliography{apssamp}

\clearpage
\onecolumngrid
\appendix
\section{Detailed numerical values}
\label{app:tables}

\setlength{\floatsep}{6pt}
\setlength{\textfloatsep}{6pt}
\setlength{\intextsep}{6pt}
\setlength{\abovecaptionskip}{4pt}
\setlength{\belowcaptionskip}{2pt}

\begin{table}[H]
\centering
\small
\setlength{\tabcolsep}{7pt}
\renewcommand{\arraystretch}{0.95}
\caption{Detailed characteristic peak positions $p^{*}$ of the response observables in two-dimensional bootstrap percolation for different thresholds $k$ and system sizes $L$. A dash ($-$) indicates that no clear extremum is observed.}
\label{tab:peak_positions}
\begin{tabular}{cccccccc}
\toprule
$k$ & $L$ & $n$ & $G_{\mathrm{site}}^{\mathrm{ns}}$ & $G_{\mathrm{edge}}^{\mathrm{ns}}$ & $B$ & $k_2(\rho)$ & $\Delta \rho$ \\
\midrule
1 & 45  & $-$ & $-$ & $-$ & $-$ & $3.50\times10^{-4}$ & $-$ \\
1 & 60  & $-$ & $-$ & $-$ & $-$ & $2.00\times10^{-4}$ & $-$ \\
1 & 75  & $-$ & $-$ & $-$ & $-$ & $1.25\times10^{-4}$ & $-$ \\
1 & 90  & $-$ & $-$ & $-$ & $-$ & $7.50\times10^{-5}$ & $-$ \\
1 & 120 & $-$ & $-$ & $-$ & $-$ & $5.30\times10^{-5}$ & $-$ \\
\midrule
2 & 30  & 0.0448 & 0.0718 & 0.0670 & 0.0540 & 0.0686 & 0.1070 \\
2 & 60  & 0.0417 & 0.0512 & 0.0496 & 0.0450 & 0.0575 & 0.0797 \\
2 & 90  & 0.0385 & 0.0449 & 0.0417 & 0.0417 & 0.0512 & 0.0702 \\
2 & 120 & 0.0385 & 0.0400 & 0.0390 & 0.0401 & 0.0496 & 0.0638 \\
\midrule
3 & 60  & $-$ & 0.5970 & 0.5830 & 0.2970 & 0.4370 & 0.5970 \\
3 & 90  & $-$ & 0.6230 & 0.6170 & 0.2970 & 0.4500 & 0.5970 \\
3 & 120 & $-$ & 0.6370 & 0.6300 & 0.2970 & 0.4430 & 0.5970 \\
3 & 180 & $-$ & 0.6700 & 0.6630 & 0.2970 & 0.4430 & 0.5970 \\
\midrule
4 & 60  & 0.1900 & 0.7630 & 0.7570 & 0.4230 & 0.5900 & 0.8030 \\
4 & 90  & 0.1900 & 0.7900 & 0.7830 & 0.4300 & 0.5830 & 0.8030 \\
4 & 120 & 0.1900 & 0.8030 & 0.7970 & 0.4300 & 0.5830 & 0.7960 \\
4 & 180 & 0.1900 & 0.8230 & 0.8170 & 0.4300 & 0.5830 & 0.8030 \\
\bottomrule
\end{tabular}
\end{table}

\vspace{-2mm}

\begin{table}[H]
\centering
\small
\setlength{\tabcolsep}{7pt}
\renewcommand{\arraystretch}{0.95}
\caption{Detailed peak amplitudes of the response observables at the corresponding characteristic positions $p^{*}$ for different thresholds $k$ and system sizes $L$. A dash ($-$) indicates that no clear extremum is observed.}
\label{tab:peak_amplitudes}
\begin{tabular}{cccccccc}
\toprule
$k$ & $L$ & $n$ & $G_{\mathrm{site}}^{\mathrm{ns}}$ & $G_{\mathrm{edge}}^{\mathrm{ns}}$ & $B$ & $k_2(\rho)$ & $\Delta \rho$ \\
\midrule
1 & 45  & $-$ & $-$ & $-$ & $-$ & 0.2499 & $-$ \\
1 & 60  & $-$ & $-$ & $-$ & $-$ & 0.2498 & $-$ \\
1 & 75  & $-$ & $-$ & $-$ & $-$ & 0.2500 & $-$ \\
1 & 90  & $-$ & $-$ & $-$ & $-$ & 0.2481 & $-$ \\
1 & 120 & $-$ & $-$ & $-$ & $-$ & 0.2494 & $-$ \\
\midrule
2 & 30  & 0.0236 & 16.4890 & 18.7790 & 0.1800 & 0.1620 & 0.8850 \\
2 & 60  & 0.0229 & 8.8397  & 12.2160 & 0.1580 & 0.1880 & 0.9180 \\
2 & 90  & 0.0224 & 5.8710  & 9.7970  & 0.1480 & 0.1963 & 0.9290 \\
2 & 120 & 0.0221 & 5.1530  & 9.2310  & 0.1430 & 0.2013 & 0.9340 \\
\midrule
3 & 60  & $-$ & 397.4430  & 450.9870  & 0.6710 & $3.93\times10^{-4}$ & 0.3040 \\
3 & 90  & $-$ & 695.9680  & 765.1510  & 0.6710 & $1.77\times10^{-4}$ & 0.3040 \\
3 & 120 & $-$ & 1024.9540 & 1096.3090 & 0.6710 & $9.80\times10^{-5}$ & 0.3040 \\
3 & 180 & $-$ & 1751.7630 & 1854.3140 & 0.6710 & $4.40\times10^{-5}$ & 0.3040 \\
\midrule
4 & 60  & 0.0796 & 978.7680  & 1223.3180 & 0.9020 & $1.14\times10^{-4}$ & 0.0820 \\
4 & 90  & 0.0796 & 1936.4850 & 2348.5940 & 0.9020 & $5.10\times10^{-5}$ & 0.0820 \\
4 & 120 & 0.0796 & 3126.7360 & 3721.7770 & 0.9020 & $2.90\times10^{-5}$ & 0.0820 \\
4 & 180 & 0.0796 & 6091.0250 & 7076.7670 & 0.9020 & $1.30\times10^{-5}$ & 0.0820 \\
\bottomrule
\end{tabular}
\end{table}

\end{document}